\documentclass[12pt]{article}

\usepackage{amssymb, amsmath}
\usepackage{graphicx}
\usepackage{cite}
\def\be{\begin{equation}}
\def\ee{\end{equation}}
\def\bea{\begin{eqnarray}}
\def\eea{\end{eqnarray}}

\def\td{ \textrm{d}}

\title{ {\bf Three-dimensional black holes and descendants}}

\author{Carmen Li\footnote{K.K.Li@sms.ed.ac.uk} \  and James Lucietti\footnote{j.lucietti@ed.ac.uk } \\ \\  \small \sl  School of Mathematics and Maxwell Institute for Mathematical Sciences, \\ \small \sl  University of Edinburgh, King's Buildings, Edinburgh, EH9 3JZ, UK}

\date{}

\begin{document}

\maketitle

\begin{abstract}
We determine the most general three-dimensional vacuum spacetime with a negative cosmological constant containing a non-singular Killing horizon. We show that the general solution with a spatially compact horizon possesses a second commuting Killing field and deduce that it must be related to the BTZ black hole (or its near-horizon geometry) by a diffeomorphism.  We show there is a general class of asymptotically AdS$_3$ extreme black holes with arbitrary charges with respect to one of the asymptotic-symmetry Virasoro algebras and vanishing charges with respect to the other. We interpret these as descendants of the extreme BTZ black hole.
 
\end{abstract}

\section{Introduction}
A central result in the theory of equilibrium black holes in four and higher dimensions is the rigidity theorem~\cite{Hawking:1971vc, Hollands:2006rj, Moncrief:2008mr}. This states that the event horizon of a stationary, rotating, black hole is a Killing horizon with respect to the Killing field
\be
K = \frac{\partial}{\partial t} + \Omega \frac{\partial}{\partial {\phi}}  \; ,   \label{K}
\ee
where $\partial /\partial t$ is the stationary Killing field, $\partial /\partial \phi$ is a Killing field generating the rotational symmetry, and $\Omega$ the angular velocity of the black hole with respect to the static asymptotic frame.\footnote{In greater than four spacetime dimensions a black hole may have multiple rotational Killing fields with corresponding angular velocities which must be included in (\ref{K}).}  For asymptotically flat space times, it is clear that if the angular velocity is non-zero, the Killing field $K$ must become spacelike outside a large enough ball.  Therefore, there is no possibility of having matter in equilibrium and co-rotating with the black hole  (since it would have to exceed the speed of light).

Black holes in anti de Sitter (AdS) spacetimes are of central importance in the context of the AdS/CFT duality~\cite{Witten:1998zw}. For such black holes the situation is quite different. In particular, for $D \geq 4$ asymptotically globally AdS black holes, such as the Kerr-AdS black hole and its higher dimensional generalisation, the Killing field $K$ is timelike everywhere outside the horizon if $| \ell \Omega | \leq 1$, where $\ell$ is the radius of AdS.  In this case $K$ defines a frame in which matter can co-rotate in equilibrium with the black hole. This raises the interesting possibility of having black holes which are invariant under a single Killing field $K$.  Matter here refers also to gravitons, hence this argument suggests the possibility of new vacuum solutions.\footnote{Such solutions would not violate the rigidity theorem since the stationary Killing field would be normal to the horizon.}  Although Kerr-AdS black holes are thought to be stable if $| \ell \Omega | \leq 1$~\cite{Hawking:1999dp}, it has been proposed that new solutions invariant under just the co-rotating Killing field may arise as the endpoint of a superradiant instability which occurs for rapidly rotating Kerr-AdS black holes (i.e. $| \ell \Omega |>1$)~\cite{Kunduri:2006qa, Cardoso:2006wa}. 
 
Finding new vacuum black hole solutions invariant under a single Killing field is a daunting task.  By coupling matter fields which are invariant under just the co-rotating Killing field, one may avoid the complication of dealing metrics with a single Killing field. Indeed, examples with a complex scalar field have been found numerically~\cite{Dias:2011at} (see also~\cite{Stotyn:2011ns, Stotyn:2013yka}). In this note we follow a different strategy by examining this problem in pure gravity in lower dimensions.

Although there are no local degrees of freedom, three-dimensional Einstein gravity with a negative cosmological constant provides a valuable toy model for examining certain higher dimensional questions~\cite{Deser:1983nh}. Brown and Henneaux demonstrated that there exist boundary conditions such that the asymptotic symmetry algebra is the infinite dimensional conformal symmetry of a cylinder~\cite{Brown:1986nw}. Furthermore, Banados-Teitelboim-Zanelli (BTZ)  found explicit black hole solutions to the $D=3$ Einstein equations~\cite{Banados:1992wn, Banados:1992gq}. Although locally AdS$_3$, globally this is a family of stationary and axisymmetric black holes which are asymptotically AdS$_3$ with a cylinder conformal boundary and possess mass $M$ and angular momentum $J$. 

The BTZ black holes always satisfy $| \ell \Omega | \leq 1$. For the non-extreme black hole ($|\ell \Omega |< 1$) the Killing field $K$ is everywhere timelike outside the horizon, whereas for the extreme black hole ($|\ell \Omega |=1$) the Killing field $K$ is everywhere null.  By the above arguments, this raises the possibility of black hole solutions invariant under a single Killing field. However, the BTZ black hole does not suffer from a superradiant instability (since it never rotates faster than the speed of light, the stability argument used in higher dimensions can be applied~\cite{Hawking:1999dp}). Therefore, such putative solutions may not arise from the evolution of some perturbation of the BTZ black hole.  Indeed, stationary and axisymmetric black holes which are coupled to a complex scalar field invariant under a co-rotating Killing field, have been argued not to exist~\cite{Stotyn:2012ap, Stotyn:2013spa}. 

In this note we show that in fact black holes with a single Killing field do {\it not} exist in three dimensional Einstein gravity, by explicitly determining the most general Einstein metric with a (non-singular) Killing horizon. It turns out the general solution with a spatially compact horizon always possesses a second commuting Killing field and hence must be related to the BTZ black hole (or its near-horizon geometry) by a diffeomorphism. Interestingly, in the case of a degenerate horizon the general solution is related to the extreme BTZ black hole by a {\it large} diffeomorphism. Our results establish a new type of uniqueness theorem for three-dimensional AdS black holes.\footnote{See e.g.~\cite{Rooman:2000ei, Fischetti:2012ps} for other types of uniqueness results.}

In fact, the general solution may have an interesting interpretation in the dual CFT. One expects that by acting on the BTZ black hole with a general element of the asymptotic-symmetry diffeomorphism group, one would obtain AdS$_3$ solutions with arbitrary Virasoro charges. We refer to these as {\it descendants} of the BTZ black hole.\footnote{Note that these are not descendants of pure states in the dual CFT.} These new solutions should also have two commuting Killing fields, corresponding to the push-forward of the Killing fields of the BTZ black hole.\footnote{We thank Harvey Reall for this observation.}  If these geometries still contain a Killing horizon, they must be within our general class of Einstein metrics.  Indeed, we identify a general class of extreme black holes that are asymptotically AdS$_3$ with cylinder boundary, which carry arbitrary charges with respect to one of the Virasoro algebras and vanishing charges with respect to the other.  Hence  these geometries are descendants (in the above sense) of the extreme BTZ black hole.

Before moving on, we mention a technical motivation which led us to investigating this problem in the extreme case.  An important inverse problem is to understand how, given a near-horizon geometry, one determines the possible corresponding extreme black holes. As we will show, three dimensional gravity provides a simple setup which allows one to examine this question explicitly.
 
\section{General solution}
\subsection{Derivation}
Consider a general $2+1$ dimensional spacetime containing a smooth\footnote{In fact, rather than smooth, we will only need to assume the functions $f,h$ are $C^1$ and $\gamma$ is $C^2$. } Killing horizon $\mathcal{N}$ of a future-pointing, complete, Killing field $K$ with a one-dimensional spacelike cross-section $H$. In the neighbourhood of $\mathcal{N}$ the metric in Gaussian null coordinates reads, see e.g.~\cite{Kunduri:2013gce}, 
\begin{equation}
\td s^2 = 2 \td v \left(  \td \lambda + \lambda h (\lambda ,x) \td x + \tfrac{1}{2} \lambda f(\lambda ,x) \td v\right)  + \gamma(\lambda, x)^2 \td x^2  \; ,   \label{horizon}
\end{equation}
where $K= \partial/ \partial v$ is the Killing field which is null on $\mathcal{N}$ and $\partial / \partial \lambda$ is tangent to null geodesics which are transverse to the horizon $\mathcal{N}$ such that $\lambda>0$ is the exterior region and $\mathcal{N}= \left\lbrace \lambda =0 \right\rbrace $. The coordinate $(x)$ is on the one-dimensional spacelike cross-section $H$, which by assumption has a non-degenerate induced metric so $\gamma>0$ in the neighbourhood of $\mathcal{N}$.  

We wish to find the general vacuum solution of this form with a cosmological constant $R_{\mu\nu} = \Lambda g_{\mu\nu}$. Of course any Einstein metric in three dimensions is {\it locally} isometric to one of the maximally symmetric spaces: we are concerned with spacetimes with a {\it global} Killing horizon as above.

To compute the Ricci tensor it is convenient to use the null-orthonormal basis $(e^{+}, e^-, e^x)$ defined by
\begin{equation}
e^+ = \td v \; ,  \qquad e^- = \td \lambda + \lambda h \td x + \tfrac{1}{2} \lambda f \td v \; , \qquad e^x = \gamma \td x  \; ,   \label{basis}
\end{equation}
so that the metric reads $\td s^2 = 2 e^+ e^- + e^x e^x$.  It turns out that the function defined by
\be
b \equiv  
 \partial_x f+\lambda f \partial_\lambda h - \lambda h \partial_\lambda f\; , \label{bdef}
\ee
appears naturally in the curvature calculations. We find that with respect to the above basis the Ricci tensor is 
\begin{eqnarray*}
&&R_{++} = \frac{1}{2 \gamma} \left[  \lambda h \partial_\lambda \left( \frac{1}{\gamma} \lambda b \right) - \partial_x \left( \frac{1}{\gamma} \lambda b \right)  - \tfrac{1}{2} \lambda^2 f^2 \partial^2_\lambda \gamma \right]  \; ,\\
&&R_{+-} = \tfrac{1}{2} \partial_\lambda^2 (\lambda f)   +\frac{1}{2 \gamma} \left[ \partial_\lambda(\lambda f \partial_\lambda \gamma) -\frac{1}{ \gamma} ( \partial_\lambda(\lambda h))^2 + \partial_x \left( \frac{1}{\gamma} \partial_\lambda(\lambda h)\right) - \lambda h \partial_\lambda \left( \frac{1}{\gamma} \partial_\lambda(\lambda h)\right) \right]  \; , \\
&&R_{+x} =  \tfrac{1}{2} \partial_\lambda \left( \frac{1}{\gamma} \lambda b \right) -\tfrac{1}{4} \lambda f \partial_\lambda \left(  \frac{1}{\gamma} \partial_\lambda (\lambda h) \right) \; ,\\
&& R_{--} = - \frac{1}{\gamma} \partial^2_\lambda \gamma \; , \qquad R_{-x} = \tfrac{1}{2} \partial_\lambda \left(  \frac{1}{\gamma} \partial_\lambda (\lambda h) \right) \; ,  \\
&&R_{xx} = \frac{1}{ \gamma} \left[ \partial_\lambda(\lambda f \partial_\lambda \gamma) -\frac{1}{2 \gamma} ( \partial_\lambda(\lambda h))^2 + \partial_x \left( \frac{1}{\gamma} \partial_\lambda(\lambda h)\right) - \lambda h \partial_\lambda \left( \frac{1}{\gamma} \partial_\lambda(\lambda h)\right) \right] \enskip .
\end{eqnarray*}
The $--$ component of the Einstein equations immediately implies $\gamma  = \gamma_0(x)+ \lambda \gamma_1(x)$, 
where $\gamma_0(x), \gamma_1(x)$ are arbitrary functions. We may use the coordinate freedom on $H$ to set $\gamma_0=1$, which we will assume henceforth.  The $-x$ component can be easily integrated for $h$ and the most general solution which is regular at $\lambda=0$ is 
\be
h = h_0(x) \left( 1+ \tfrac{1}{2} \lambda \gamma_1(x) \right),
\ee where $h_0$ is an arbitrary function. Now consider the ${+-}$ and ${xx}$ components. This is facilitated by noting that the Einstein equation implies  $\tfrac{1}{2}R_{xx} - R_{+-}=- \tfrac{1}{2}\Lambda$, which explicitly reads
\be
\partial_\lambda^2 (\lambda f) -\tfrac{1}{2} \left( \frac{ \partial_\lambda (\lambda h)}{\gamma} \right)^2 = \Lambda  \; .
\ee
This can now be integrated for $f$ and the most general solution regular at $\lambda=0$ is
\be
f = f_0(x) + \tfrac{1}{2}\left( \Lambda + \tfrac{1}{2} h_0(x)^2 \right) \lambda   \label{f}
\ee
where $f_0$ is an arbitrary function. Now, the $xx$ equation is satisfied iff
\be
\partial_x h_0 - \tfrac{1}{2} h_0^2 + f_0 \gamma_1 = \Lambda  \; .
\ee
It remains to consider the $++$ and $+x$ components. It is easy to see these are satisfied if and only if $\lambda b / \gamma$ is a constant. Hence by regularity at $\lambda=0$ we deduce $b=0$. Finally, by substituting into (\ref{bdef}) one finds $b = \partial_x f_0$ and so we deduce that $f_0(x) = - 2\kappa$, where $\kappa$ is a constant. We have now satisfied all components of the Einstein equation.

To summarise, we have found that the most general solution with a non-singular Killing horizon is given by:
\bea
\gamma(\lambda, x)  &=& 1 + \lambda \gamma_1(x)  \nonumber  \\ 
h(\lambda, x) &=& h_0(x) \left( 1+ \tfrac{1}{2} \lambda \gamma_1(x) \right)  \nonumber  \\
f(\lambda, x) &=& -2\kappa + \tfrac{1}{2}\lambda \left( \Lambda + \tfrac{1}{2} h_0(x)^2 \right)  \; ,  \label{soln}
\eea
where $\kappa$ is a constant and $h_0, \gamma_1$ are arbitrary functions subject to the constraint
\be
\partial_x h_0 - \tfrac{1}{2} h_0^2 -2\kappa \gamma_1 = \Lambda  \; .   \label{constr}
\ee
The various quantities which appear in the solution all have a direct geometrical meaning.  The 1-form $h_0\td x$ is the connection of the normal bundle on $H$, viewed as a submanifold of the spacetime. The function $\gamma_1= \theta|_{\lambda=0}$ where  $\theta$ is the expansion of the null geodesic congruence tangent to $\partial / \partial \lambda$, i.e. $\theta=\gamma^{-1}\partial_\lambda \gamma$. The constant $\kappa$ is the surface gravity on the Killing horizon, i.e. $\td K^2 |_{\lambda=0} = - 2\kappa K|_{\lambda=0}$.

In the non-degenerate case, $\kappa \neq 0$, the constraint equation (\ref{constr}) can be solved to determine the extrinsic data $\gamma_1$ in terms of the intrinsic data $h_0$, so the solution depends on the constant $\kappa$ and one freely specifiable function on the horizon $h_0(x)$.  On the other hand, in the degenerate case, $\kappa=0$, we deduce that the constraint equation (\ref{constr}) reduces to the Einstein equation for the near-horizon geometry~\cite{Kunduri:2013gce} and $\gamma_1(x)$ is an {\it arbitrary} function on $H$. Hence, once the near-horizon solution has been fixed, the degenerate solution depends only on one freely specifiable function $\gamma_1(x)$. This explicitly shows that decoupling of intrinsic and extrinsic data occur if and only if the horizon is degenerate. 

In general, Gaussian null coordinates are only defined in a neighbourhood on the horizon, in particular, as long as the transverse null geodesic congruence $\partial /\partial \lambda$ does not caustic. For our solution,   observe that if $\gamma_1(x_0)<0$ for some $x_0$ the transverse null geodesics converge initially, i.e. $\theta(\lambda, x_0)<0$ for small $\lambda$,  and furthermore $\theta \to -\infty$ as $\lambda \to 1/|\gamma_1(x_0)|$. On the other hand, if $\gamma_1(x)\geq 0$ it is clear the coordinate system can be extended to all positive values of $\lambda$.

We emphasise that our general solution is valid for any cosmological constant. Motivated by the discussion in the introduction, in this note we will focus on AdS solutions with compact cross-sections of the horizon. Therefore, henceforth we set $\Lambda = - 2/ \ell^2$ and $H \cong S^1$. We thus identify $x\sim x +2\pi R$, where $R>0$ is the radius of the horizon, and assume the functions $h_0(x), \gamma_1(x)$ are $2\pi R$-periodic.  Thus, if $\gamma_1(x)>0$, then $\lambda \to \infty$ is a conformal boundary with boundary metric
\be
\lambda^{-2} \td s^2 \to - \frac{\td v^2}{\ell^2} + (\gamma_1 \td x+ \tfrac{1}{2}h_0 \td v)^2  \; .   \label{bdy}
\ee
If $h_0$ is constant we may define coordinates $t=v$ and $\td \phi =\gamma_1(x) \td x + \tfrac{1}{2}h_0 \td v$ which explicitly show the boundary is a flat cylinder. This will be relevant below.

\subsection{Extra Killing field}

 We will now show that under the assumptions $H \cong S^1$ and $\Lambda<0$, our general solution in fact always possesses a second Killing field which commutes with $K$ and is globally defined (i.e. it is compatible with the periodic identification $x \sim x+2\pi R$).

A tedious calculation shows that for the general non-degenerate case $\kappa \neq 0$, the most general Killing field which commutes with $K$ is (a multiple of)
\be
X = \left( c+ \frac{h_0}{2\kappa} \right) \partial_ v + \frac{\lambda^2 h_0 h_0'}{4\kappa \gamma}  \partial_\lambda +  \left( 1 - \frac{\lambda h_0'}{2\kappa \gamma} \right)   \partial_x  \; ,
\ee
where $c$ is a constant and we have used (\ref{constr}). Observe that this Killing field is globally defined, tangent to the horizon $\mathcal{N}$ and has closed orbits. 

For the general degenerate case $\kappa=0$,  equation (\ref{constr}) shows $h_0$ is determined by the near-horizon equation
\be
\partial_x h_0 - \tfrac{1}{2} h_0^2  =  - \frac{2}{\ell^2}  \; . \label{nh}
\ee
It has been shown that the most general solution on $H \cong S^1$  is $h_0 = 2 / \ell$ (choosing a sign), which corresponds to the near-horizon geometry of the extreme BTZ black hole~\cite{Kunduri:2013gce}. In this case, it can be shown that the most general globally defined Killing field which commutes with $K$ is (a multiple of)
\be
X  =(c+ y) \partial_ v +\frac{\lambda^2 y'}{\ell \gamma} \partial_\lambda + \left(1 - \frac{\lambda y'}{\gamma} \right)  \partial_x  \; ,   \label{Xex}
\ee
where $c$ is a constant and $y(x)$ is the unique periodic solution to
\be
y' - \frac{2}{\ell} y =  \gamma_1(x)  \; .  \label{yeq}
\ee
Again, note that this Killing field has closed orbits and is also tangent to the horizon.

Thus in either case we see that a general spacetime containing a Killing horizon with compact cross-sections always possesses a second Killing field $X$ with closed orbits which commutes with $K$, i.e. it is {\it axisymmetric}.\footnote{We emphasise that, although related, this does not follow from the usual rigidity theorem for stationary rotating black holes.} Since $X$ is tangent to the horizon, we could always choose a different cross-section $\tilde{H} \cong S^1$ such that $X$ is tangent to $\tilde{H}$ for some constant $c$.  In this case, the solution written in Gaussian null coordinates $(\tilde{v}, \tilde{\lambda}, \tilde{x}$) adapted to this new cross-section $\tilde{H}$, must take our general form  (\ref{soln}) but with $\tilde{h}_0, \tilde{\gamma}_1$ constant functions. It is then easy to see the solution is given by the BTZ black hole or its near-horizon geometry, as we show next. 

Thus suppose that $\partial /\partial x$ is a Killing field with closed orbits so $h_0$ and $\gamma_1$ are constant functions. Using the discrete transformations $x \to -x$  and $(v,\lambda, \kappa) \to -(v, \lambda, \kappa)$  we may always arrange  $h_0\geq 0$ and $\gamma_1\geq 0$, respectively. 

If $\gamma_1 > 0$ define two positive parameters $(r_+, r_-)$ by $\gamma_1=1/r_+$ and  $h_0 =  2r_-/(\ell r_+)$. Solving the constraint (\ref{constr}) implies $\kappa  = (r_+^2- r_-^2)/(\ell^2 r_+)$. Now, performing the coordinate change
\bea
\lambda &=&r-r_+\;, \nonumber \\  \td v &=& \td t + \frac{\td r}{N^2}  \nonumber  \\ \td x &=& r_+ \td \phi - \frac{r_-}{\ell} \td t+r_+ \left( N^\phi - \frac{r_-}{r_+\ell} \right) \frac{\td r}{N^2}  \; , \label{btzcoords}
\eea
where we have defined the functions
\be
N^2 = \frac{(r^2-r_+^2)(r^2-r_-^2)}{\ell^2 r^2} \; , \qquad \qquad {N}^\phi =  \frac{r_- r_+}{\ell r^2}   \; ,
\ee
gives
\be
\td s_{\text{BTZ}}^2 = - N^2 \td t^2 + \frac{\td r^2}{N^2} + r^2 \left( \td {\phi} + N^\phi \td t \right)^2  \; ,
\ee
which is the BTZ black hole solution.  If $\kappa\geq 0$ the horizon $\lambda=0$ corresponds to the outer horizon, whereas if $\kappa<0$ the horizon $\lambda=0$ corresponds to the inner horizon.

If  $\gamma_1=0$, the constraint (\ref{constr}) can be immediately solved to get  $h_0 = 2/\ell$ and hence the solution in this case simply reads
\be
\td s^2 = \left( -2 \kappa \lambda - \frac{\lambda^2}{\ell^2} \right) \td v^2 + 2 \td v \td \lambda + \left( \td x + \frac{\lambda \td v}{\ell} \right)^2   \; .
\ee
If $\kappa=0$ this is the near-horizon limit of the extreme BTZ black hole. If $\kappa \neq  0$ this is the decoupling limit of the near-extreme BTZ black hole. 

\section{General solution with a degenerate horizon}

In this section we will study the general solution containing a degenerate horizon ($\kappa=0$) with compact cross-sections $H \cong S^1$. A shown above, the general spacetime in this case is given by 
\be
\td s^2 = 2 \td v \left[ \td \lambda + \frac{2}{\ell}  \lambda(1+\tfrac{1}{2} \lambda \gamma_1(x))\td x \right] + (1+\lambda \gamma_1(x))^2 \td x^2   \; , \label{deg}
\ee
where $\gamma_1(x)$ is an arbitrary periodic function $\gamma_1(x+2\pi R) =\gamma_1(x)$. 

\subsection{Large diffeomorphism}
We now explicitly show that this solution is globally isometric to the BTZ black hole, or its near-horizon geometry, by introducing coordinates adapted to the two commuting Killing fields $K=\partial/\partial v$ and $X$ given by (\ref{Xex}).

The inner products of the Killing fields are thus:
\bea
K^2=0, \qquad K \cdot X = \frac{2\lambda}{\ell} \left( 1- \frac{\lambda y}{\ell} \right), \qquad X^2 = 1+ \frac{4c\lambda}{\ell} \left( 1- \frac{\lambda y}{\ell} \right)  \; .
\eea
Define a third vector field $U$ by: $U^2=0, U \cdot X=0, U \cdot K=1$. It is easy to show that
\be
U = -\frac{1}{2} C^2 \partial_v+ \partial_\lambda +C e_x   \label{Uex}
\ee
where $e_x= \frac{1}{\gamma}( \partial_x - \lambda h \partial_\lambda)$ is a dual vector to the basis (\ref{basis}) and the function $C$ satisfies
\be
y+c + \left( 1- \frac{2 \lambda y}{\ell} \right) C - \frac{\lambda}{\ell} \left(1-\frac{\lambda y}{\ell} \right) C^2=0   \; . 
\ee
The discriminant of this quadratic is simply $X^2$. Hence the unique solution which is regular on the horizon is
\be
C = \frac{ 1- \frac{2\lambda y}{\ell} - \sqrt{1+ \frac{4c\lambda}{\ell} \left( 1- \frac{\lambda y}{\ell} \right)}}{\frac{2}{\ell} \lambda (1 - \frac{\lambda y}{\ell})}  \; , \label{Ceq}
\ee
where we must have $X^2>0$.
Now a tedious calculation shows that $[X, U]=0$ if and only if
\be
\frac{\lambda^2 y'}{\ell} \partial_\lambda C + \left(1 - \frac{2\lambda y}{\ell} \right) \partial_x C + y'=0  \; . \label{XU}
\ee
Remarkably, it can be shown that (\ref{Ceq}) automatically satisfies (\ref{XU}). This allows us to deduce that a new coordinate system $(\tilde{v}, \tilde{\lambda}, \tilde{x})$ exists such that
\be
K = \frac{\partial}{ \partial \tilde{v}}, \qquad U = \frac{\partial}{\partial \tilde{\lambda}}, \qquad X= \frac{\partial}{\partial \tilde{x}}   \; .
\ee
From (\ref{Uex}) we may read off  $ \frac{\partial \lambda}{\partial \tilde{\lambda}} = 1- \frac{\lambda h C}{\gamma} $ and $\frac{\partial x}{\partial \tilde{\lambda}} = \frac{C}{\gamma}$ which imply
\bea
\partial_{\tilde{\lambda}}  \sqrt{X^2} &=&\frac{2c}{\ell} \frac{\left[ 1- \frac{2\lambda y}{\ell} - \frac{2\lambda}{\ell} \left( 1- \frac{\lambda y}{\ell} \right) C \right] }{\sqrt{1+ \frac{4c\lambda}{\ell} \left( 1- \frac{\lambda y}{\ell} \right) }} = \frac{2c}{\ell}  \; ,
\eea
where in the second equality we used (\ref{Ceq}). Hence, integrating and fixing the horizon to be at $\tilde{\lambda}=0$ we get
\be
\sqrt{X^2}= 1+\frac{2c \tilde{\lambda}}{\ell}  \;, \qquad K \cdot X =  \frac{2\tilde{\lambda}}{\ell} \left( 1+ \frac{c \tilde{\lambda}}{\ell} \right)   \; .
\ee
Therefore, the metric in the new coordinates is
\be
\td s^2 = 2 \td \tilde{v} \left[ \td \tilde{\lambda} + \frac{2}{\ell} \tilde{\lambda}\left(1+\frac{c \tilde{\lambda}}{\ell} \right)\td \tilde{x} \right] + \left(1+\frac{2 c \tilde{\lambda}}{\ell} \right)^2 \td \tilde{x}^2   \; .
\ee
This expresses the solution in Gaussian null coordinates adapted to a cross-section $\tilde{H} \cong S^1$ which is tangent to $X$. It is thus takes our general form (\ref{deg}) with $\tilde{\gamma_1} = 2c /\ell$ a constant. As we showed above this is the extreme BTZ black hole ($c \neq 0$) or its near-horizon geometry $(c=0)$.

The results of the next section will show that the diffeomorphism constructed above must be a large diffeomorphism.

\subsection{Asymptotic charges}

We now consider the extreme solution in a chart adapted to a general cross-section, i.e. the spacetime (\ref{deg}). We will assume that the transverse null geodesics $\partial /\partial \lambda$ are strictly expanding, i.e. $\gamma_1(x)>0$.  This ensures it is asymptotically AdS$_3$ with a cylinder conformal boundary.  In fact since $h_0$ is constant and $x$ is periodically identified, this immediately follows from (\ref{bdy}).

To see this in more detail, consider the coordinate change defined by
\bea
r &=& \lambda + \frac{1}{\gamma_1(x)}  \nonumber \\
 t &=& v + \frac{\ell^2}{r} \left(1+ \frac{\beta(x)}{3 r^2} \right) \nonumber \\
 \phi &=& \int \gamma_1(x) \td x +  \frac{v}{\ell}   - \frac{\beta(x)}{3r^3}  \label{ads3coords}
 \eea 
 where the function
 \be
 \beta \equiv  \frac{1}{\gamma_1^2} \left( 1 - \frac{\ell \gamma_1'}{\gamma_1} \right)  \;   .  \label{beta}
 \ee
To derive this coordinate change, we expanded the one for extreme BTZ (\ref{btzcoords}) for large $r$ and then allowed the subleading terms to depend on $x$. 
 Observe that the coordinate change (\ref{ads3coords}) forces $\phi$ to be a periodic coordinate with period $\int_0^{2\pi R} \gamma_1(x) \td x$.  By scaling $(v, \lambda, \gamma_1) \to ( c v, c^{-1} \lambda, c\gamma_1)$, we may always fix the period of $\phi$ to be $2\pi$. In these coordinates our general metric (\ref{deg}) has the following asymptotics
\bea
 g_{tt} &=& - \frac{r^2}{\ell^2} + \frac{2 \beta(x)}{\ell^2}+ \mathcal{O} (r^{-1}) \;, \qquad g_{t \phi} =  -\frac{\beta(x)}{\ell} + \mathcal{O}(r^{-1})  \; , \qquad  g_{\phi \phi} = r^2 + \mathcal{O}(r^{-1}) \; ,\nonumber \\
g_{tr} &=& - \frac{2 \ell \beta'(x)}{3 \gamma_1(x) r^3} + \mathcal{O}(r^{-4}) \; , \qquad g_{r \phi} = \frac{\ell^2 \beta'(x)}{3 \gamma_1(x) r^3} +\mathcal{O}(r^{-4}) \; , \nonumber  \\ g_{rr} &=& \frac{\ell^2}{r^2} + \frac{2\ell^2 \beta(x)}{r^4} +\mathcal{O}(r^{-5}) \; ,   \label{asymptoticmetric}
\end{eqnarray}
for $r \to \infty$, which  explicitly shows that our spacetime is asymptotically AdS$_3$ in the sense of Brown and Henneaux~\cite{Brown:1986nw}.  
Observe that for large $r$
\be
\phi - \frac{t}{\ell}  = \int \gamma_1(x) \td x   + \mathcal{O}(r^{-1})
\ee
and hence asymptotically $x$ is purely a function of $\phi - \frac{t}{\ell}$ (note the coordinate change is invertible due to our assumption $\gamma_1>0$). 

From the subleading terms in (\ref{asymptoticmetric}) we may compute the asymptotic charges of this solution.  The asymptotic-symmetry generators  are~\cite{Brown:1986nw}
\be
L^\pm_n = \frac{1}{2} e^{ i n ( \frac{t}{\ell} \mp \phi ) } \left(\ell \frac{\partial}{\partial t} \mp  \frac{\partial}{\partial \phi}  \right)  + \dots
\ee
where $\dots$ denotes subleading terms and also terms proportional to $\partial_r$ which will not be needed.  The conserved charge $Q[\xi]$ associated to an asymptotic symmetry generated by a vector field $\xi$ is an integral at fixed time $t$ over the boundary circle at spacelike infinity $r\to \infty$.  We find that, relative to the zero mass BTZ solution, the Virasoro charges are
\bea
Q[L^+_n] &=& \frac{1}{\ell \pi} \int_0^{2\pi} \td \phi  \;  e^{ i n ( \frac{t}{\ell} - \phi ) }  \beta(x)   \\
Q[L^-_n] &=& 0 \; ,
\eea
where as noted above asymptotically $x$ is only a function of $\phi - \frac{t}{\ell}$.  Thus our general solution generically carries non-zero charges only in one of the Virasoro algebras.
In particular, the mass $\ell M = Q[L_0^+]+Q[L_0^-]$ and angular momentum $J = Q[L_0^+]-Q[L_0^-]$  are given by
\be
\ell M = J =  \frac{1}{\ell \pi} \int_0^{2\pi} \td \phi  \;   \beta(x)   =  \frac{1}{\ell \pi} \int_0^{2\pi R}  \frac{\td x }{\gamma_1(x)} \; ,
\ee
where in the final equality we converted to the $x$ coordinate (at constant $t$) and used the explicit form of $\beta$ together with periodicity.

Thus we see that the mass/angular-momentum relation satisfied by the extreme BTZ black hole persists for this class of spacetimes.  However, unlike the BTZ black hole, these carry arbitrary non-zero charges with respect to all the Virasoro generators $L^+_n$ and vanishing ones with respect to $L^-_n$. In particular, the general solution is characterised by the Virasoro charges $Q[L^+_n]$ with $n \neq 0$. It is worth noting that if $Q[L^+_n]=0$ for all $n \neq 0$, then the function $\beta$ must be a constant and hence (\ref{beta}) implies $\gamma_1$ must be a constant (using periodicity) and we recover the BTZ black hole. Therefore, these geometries may be interpreted as descendants of the extreme BTZ black hole.

\section{Non-degenerate horizon}

In this section we study the general solution containing a non-degenerate horizon ($\kappa \neq 0$) with compact cross-sections $H \cong S^1$. As shown above, the general solution is given by (\ref{horizon}), (\ref{soln}) and is determined by the constant $\kappa$ and an arbitrary function $h_0(x)$, with $\gamma_1(x)$ then determined by (\ref{constr}). These functions must be periodic with $x\sim x+2\pi R$.

We will also assume that $\gamma_1(x)>0$ so the transverse null geodesics $\partial /\partial \lambda$ are strictly expanding and that $\kappa>0$ to ensure the null generators are future complete.  Under these conditions, it can be shown that (\ref{constr}) implies
\be
h_0(x)^2 < \frac{4}{\ell^2}  \; ,  \label{h0ineq}
\ee
for {\it all} $x$ (otherwise $h_0$  is monotonic, contradicting periodicity). In fact, this condition implies the Killing field $K$ is timelike {\it everywhere} outside the horizon.

We will first analyse the conformal boundary of this Einstein spacetime. The main complication arises due to the fact that the conformal boundary metric in the frame defined by equation (\ref{bdy}) is not flat for non-constant $h_0(x)$. Remarkably, we find there is a simple Weyl transformation on the boundary which makes (\ref{bdy}) a flat cylinder and is consistent with the global identifications we already have (i.e. $x$ periodic and $v$ not). It may be verified that
\be
\td s^2_b= \frac{- \frac{\td v^2}{\ell^2} + (\gamma_1 \td x+ \tfrac{1}{2}h_0 \td v)^2}{\Omega(x)^2}  \; ,    \label{bdy2}
\ee
where 
\be
\Omega(x)  \equiv \sqrt{1- \frac{\ell^2 h_0^2}{4} } \;  ,
\ee
is a flat metric for any $h_0(x), \gamma_1(x)$. Indeed, this can be seen by performing the coordinate change
\be
\td t = c_\beta \td v + \frac{\ell \gamma_1}{\Omega^2} \left( s_\beta -  \frac{c_\beta \ell h_0}{2} \right) \td x \; ,\qquad \td \phi =  s_\beta \frac{\td v}{\ell}+  \frac{\gamma_1}{\Omega^2}\left( c_\beta -  \frac{s_\beta \ell h_0}{2} \right)  \td x  \; ,
\ee
where $c_\beta = \cosh \beta, s_\beta = \sinh \beta$ and $\beta$ is a constant ``boost" parameter,
which gives 
\be
\td s_b^2 = -\frac{\td t^2}{\ell^2} + \td \phi^2  \; .
\ee
Observe that we have to include the boost $\beta$ (which is a large diffeomorphism) in order to ensure that the coordinate $t$ is not periodically identified. The condition to avoid identifications of $t$ corresponds to the following specific choice of boost:
\be
\tanh \beta =\frac{ \int_{0}^{2\pi R}  \frac{\ell h_0 \gamma_1}{2\Omega^2} \td x}{ \int_0^{2 \pi R} \frac{\gamma_1}{\Omega^2} \td x}  \; ,
\ee
which we assume henceforth. Observe that due to $\gamma_1>0$ and (\ref{h0ineq}), 
\be
\left| \frac{ \int_{0}^{2\pi R}  \frac{\ell h_0 \gamma_1}{2\Omega^2} \td x}{ \int_{0}^{2\pi R} \frac{\gamma_1}{\Omega^2} \td x} \right| \leq \frac{ \int_{0}^{2\pi R} \frac{\ell |h_0| \gamma_1}{2\Omega^2} \td x}{\int_{0}^{2\pi R} \frac{\gamma_1}{\Omega^2} \td x} <1 ,
\ee
so a unique value for this special  boost always exists. Furthermore, by a discrete transformation $x \to -x$ we may always arrange $\beta \geq 0$, which we will assume below. Also note that $\partial \phi/ \partial x >0$ so the coordinate $\phi$ inherits a periodicity from $x$.  By scaling $(v, \lambda, \gamma_1) \to ( c v, c^{-1} \lambda, c\gamma_1)$, we may always fix the period of $\phi$ to be $2\pi$.   Hence, in the above conformal frame the boundary is indeed globally a flat cylinder, as claimed. 

We will now compute the asymptotic charges by working in the cylinder conformal frame. Consider the coordinate change defined by
\bea
r &=& \Omega \left( \lambda + \frac{\ell^2 \kappa}{\Omega^2} \right) \; ,\nonumber \\
t &=& c_\beta v  + \int \frac{\ell \gamma_1}{\Omega^2}\left( s_\beta - \frac{c_\beta \ell h_0}{2} \right) \td x   + \frac{\ell^2}{\Omega r}\left[ c_\beta - \frac{s_\beta \ell h_0}{2}+ \frac{\kappa^2\ell^4}{3\Omega^2 r^2}\left( c_\beta - \frac{s_\beta \ell^3 h_0^3}{8} \right) \right]  \; ,\nonumber \\
\phi &=& s_\beta \frac{v}{\ell}+ \int  \frac{\gamma_1}{\Omega^2}\left( c_\beta - \frac{s_\beta \ell h_0}{2} \right) \td x   + \frac{\ell}{\Omega r} \left[ s_\beta - \frac{c_\beta \ell h_0}{2} +\frac{\kappa^2 \ell^4}{3\Omega^2 r^2} \left( s_\beta - \frac{c_\beta \ell^3 h_0^3}{8} \right) \right] \; .  \label{phigen}
\eea
In these coordinates we find that the metric for $r \to \infty$ has the following behaviour:
\bea
&&g_{tt} = -\frac{r^2}{\ell^2} + \frac{\kappa^2\ell^2}{\Omega^2}\left( c_\beta^2 - \frac{s_\beta^2 \ell^2 h_0^2}{4} \right) +\mathcal{O}(r^{-3}) \; , \qquad \quad g_{t\phi} =  -s_\beta c_\beta \kappa^2 \ell^3 + \mathcal{O}(r^{-3}) \; , \nonumber \\ && g_{\phi\phi} = r^2  + \frac{\kappa^2 \ell^4}{\Omega^2}\left( s_\beta^2 - \frac{c_\beta^2 \ell^2 h_0^2}{4} \right)+ \mathcal{O}
(r^{-3})  \nonumber \; , \qquad \qquad g_{tr}= \mathcal{O}(r^{-3}) \; , \\  && g_{\phi r} = \mathcal{O}(r^{-3})  \; ,\qquad \qquad g_{rr} = \frac{\ell^2}{r^2}+ \frac{\kappa^2 \ell^6}{\Omega^2 r^4} \left(1+ \frac{\ell^2 h_0^2}{4} \right) +\mathcal{O}(r^{-5})  \; ,
\eea
which is of Brown-Henneaux form and hence suitable for reading off the conserved charges.  Observe that asymptotically $x$ is a function of {\it both} $\phi \pm t/\ell$. Hence a priori, one would expect the Virasoro charges $Q[L^\pm_n] \neq 0$ for all integers $n$. In fact, performing the computation, we actually find that
\be
Q[L_n^\pm] = \ell^3 \kappa^2 \left( \frac{1}{2} +s_\beta^2 \pm s_\beta c_\beta \right) \delta_{n,0}  \; .
\ee
Therefore, we see that all higher Virasoro charges vanish, whereas the zero mode charges give the following mass and angular momentum
\be
M = \ell^2 \kappa^2 ( c_\beta^2+ s_\beta^2), \qquad \qquad J = 2\ell^3 \kappa^2 s_\beta c_\beta   \; .
\ee
Observe that
\be
\ell M - J = \ell^3 \kappa^2 (c_\beta -  s_\beta )^2>0.
\ee
Therefore, our spacetime has precisely the same Virasoro charges as the non-extreme BTZ black hole. It follows that it must be diffeomorphic to the non-extreme BTZ black hole.\footnote{This follows from the fact that the Fefferman-Graham expansion (\ref{FG}) terminates in three-dimensions and that $Q[L_n^\pm] \sim \int_0^{2\pi} \td \phi \; e^{in x^\pm} T^\pm(x^\pm) $.} Hence, in contrast to the extreme case, we do not obtain descendants of the non-extreme BTZ black hole (which would possess arbitrary charges with respect to all $L^\pm_n$ and thus be related by a large diffeomorphism).
\section{Discussion}

Another way of understanding our results may be as follows. The Fefferman-Graham expansion for three-dimensional Einstein spacetimes terminates and hence the conformal boundary metric and stress tensor determine the full spacetime~\cite{Skenderis:1999nb}.  For asymptotically globally AdS$_3$ spacetimes, so a cylinder conformal boundary metric $-\frac{\td t^2}{\ell^2}+ \td \phi^2$ where $\phi \sim \phi +2\pi$,  it is easy to determine the general Einstein spacetime~\cite{Banados:1998gg}, which is
\be
\td s^2  = \frac{1}{z^2} \left[\ell^2 \td z^2  + 2 ( \td x^+ + \tfrac{1}{2}z^2{T^-}(x^-) \td x^- ) ( \td x^- +\tfrac{1}{2} z^2{T^+}(x^+) \td x^+ ) \right]  \; , \label{FG}
\ee
where $z=0$ is the conformal boundary, $\sqrt{2} x^\pm =\phi \pm \frac{t}{\ell}$ are lightcone coordinates on the cylinder, and the two arbitrary functions $T^\pm(x^\pm)$ are the components of the boundary stress tensor.  

Now, suppose (\ref{FG}) describes a black hole with a horizon invariant under a Killing field of the form (\ref{K}). If $|\ell \Omega | \neq 1$, then it is straightforward to show that {\it both} $T^\pm(x^\pm)$ must be constant functions and hence the spacetime is stationary and axisymmetric (if $T^\pm >0$ this is the BTZ black hole). On the other hand, if $|\ell \Omega| =1$, then only one of $\partial_+$ or $\partial_-$ is a Killing field; without loss of generality suppose $\Omega \ell=1$ so $K\propto \partial_+$. Then $T^+(x^+)$ is again a constant, although now $T^-(x^-)$ can be an arbitrary function. It then follows that $|K|^2 = T^+$ is a constant and therefore such a Killing horizon exists if and only if $T^+=0$.  In this case, $K$ is a globally null Killing field and since by assumption it is tangent to the black hole horizon, the horizon must be {\it degenerate}.  

This simple argument allows for extreme black holes more general than extreme BTZ, which are related by a large diffeomorphism to the extreme BTZ. Furthermore, it also does not appear to allow for more general non-extreme black holes with a {\it Killing} horizon. This picture is consistent with the results derived in this paper, which were obtained by determining the most general three-dimensional Einstein metric containing a Killing horizon. 

The asymptotic Virasoro charges of our general extreme black hole show that these geometries can be interpreted as descendants of the extreme BTZ (as defined in the introduction).  It would be interesting to better understand their CFT interpretation. 
On the other hand, we did not find descendants of the non-extreme BTZ black hole within our general solution with a non-extreme horizon (these would be related by a large diffeomorphism).\footnote{Our only assumption was that the transverse null geodesics are strictly expanding, so $\gamma_1(x)>0$ for all $x$. It would be interesting to analyse the other cases to confirm the generality of this statement.} This indicates that the descendants of the non-extreme BTZ black hole do {\it not} have a Killing horizon. It would be interesting to understand this by directly analysing under what conditions the general Einstein metric (\ref{FG}) contains a non-singular horizon. \\

\noindent {\bf Acknowledgements}. We would like to thank Mukund Rangamani, Harvey Reall and Joan Simon for useful comments. We would especially like to thank Don Marolf for pointing out the possible existence of a second Killing field, as well as Harvey Reall and Joan Simon for further insights. CL is supported by a Principal Career Development Scholarship at the University of Edinburgh. JL is supported by an EPSRC Career Acceleration Fellowship.


\begin{thebibliography}{99}
{  \small

\bibitem{Hawking:1971vc}
  S.~W.~Hawking,
  Commun.\ Math.\ Phys.\  {\bf 25} (1972) 152.
  
\bibitem{Hollands:2006rj}
  S.~Hollands, A.~Ishibashi and R.~M.~Wald,
  Commun.\ Math.\ Phys.\  {\bf 271} (2007) 699
  [gr-qc/0605106].
  
  \bibitem{Moncrief:2008mr}
  V.~Moncrief and J.~Isenberg,
  Class.\ Quant.\ Grav.\  {\bf 25} (2008) 195015
  [arXiv:0805.1451 [gr-qc]].
  
\bibitem{Witten:1998zw}
  E.~Witten,
  Adv.\ Theor.\ Math.\ Phys.\  {\bf 2} (1998) 505
  [hep-th/9803131].

\bibitem{Hawking:1999dp}
  S.~W.~Hawking and H.~S.~Reall,
  Phys.\ Rev.\ D {\bf 61} (2000) 024014
  [hep-th/9908109].

\bibitem{Kunduri:2006qa}
  H.~K.~Kunduri, J.~Lucietti and H.~S.~Reall,
  Phys.\ Rev.\ D {\bf 74} (2006) 084021
  [hep-th/0606076].
  
  \bibitem{Cardoso:2006wa}
  V.~Cardoso, O.~J.~C.~Dias and S.~Yoshida,
  Phys.\ Rev.\ D {\bf 74} (2006) 044008
  [hep-th/0607162].

\bibitem{Dias:2011at}
  O.~J.~C.~Dias, G.~T.~Horowitz and J.~E.~Santos,
  JHEP {\bf 1107} (2011) 115
  [arXiv:1105.4167].
  
  
\bibitem{Stotyn:2011ns}
  S.~Stotyn, M.~Park, P.~McGrath and R.~B.~Mann,
  Phys.\ Rev.\ D {\bf 85} (2012) 044036
  
\bibitem{Stotyn:2013yka}
  S.~Stotyn, C.~D.~Leonard, M.~Oltean, L.~J.~Henderson and R.~B.~Mann,
  arXiv:1307.8159 [hep-th].
  
  
\bibitem{Deser:1983nh}
  S.~Deser and R.~Jackiw,
  Annals Phys.\  {\bf 153} (1984) 405.

\bibitem{Brown:1986nw}
  J.~D.~Brown and M.~Henneaux,
  Commun.\ Math.\ Phys.\  {\bf 104} (1986) 207.


\bibitem{Banados:1992wn}
  M.~Banados, C.~Teitelboim and J.~Zanelli,
  Phys.\ Rev.\ Lett.\  {\bf 69} (1992) 1849
  [hep-th/9204099].
  
\bibitem{Banados:1992gq}
  M.~Banados, M.~Henneaux, C.~Teitelboim and J.~Zanelli,
  Phys.\ Rev.\ D {\bf 48} (1993) 1506
  [gr-qc/9302012].
  
\bibitem{Stotyn:2012ap}
  S.~Stotyn and R.~B.~Mann,
  J.\ Phys.\ A {\bf 45} (2012) 374025
  [arXiv:1203.0214 [gr-qc]].
  
\bibitem{Stotyn:2013spa}
  S.~Stotyn, M.~Chanona and R.~B.~Mann,
  arXiv:1309.2911 [hep-th].
  
\bibitem{Rooman:2000ei}
  M.~Rooman and P.~.Spindel,
  Class.\ Quant.\ Grav.\  {\bf 18} (2001) 2117
  [gr-qc/0011005].
  
  \bibitem{Fischetti:2012ps}
  S.~Fischetti and D.~Marolf,
  Class.\ Quant.\ Grav.\  {\bf 29} (2012) 105004
  [arXiv:1202.5069 [hep-th]].
  
  \bibitem{Skenderis:1999nb}
  K.~Skenderis and S.~N.~Solodukhin,
  Phys.\ Lett.\ B {\bf 472} (2000) 316
  [hep-th/9910023].
  
  \bibitem{Banados:1998gg}
  M.~Banados,
  hep-th/9901148.
  
\bibitem{Kunduri:2013gce}
  H.~K.~Kunduri and J.~Lucietti,
  Living Rev.\  Rel.\  {\bf 16} (2013) 8
  [arXiv:1306.2517 [hep-th]].


}
\end{thebibliography}
\end{document}